\input harvmac.tex

\def\a{{\alpha}}
\def\b{\beta}

\def\p{{\partial}}
\def\e{{\epsilon}}
\def\wdg{{\wedge}}

\def\frac#1#2{{#1\over #2}}
\def\half{{ {\frac{1}{2}} }}

\lref\DRS{K. Dasgupta, G. Rajesh, S. Sethi,
``M Theory, Orientifolds and G-Flux'', 
hep-th/9908088, JHEP 9908 (1999) 023.}
\lref\BD{K. Becker, K. Dasgupta, ``Heterotic Strings with Torsion'',
hep-th/0209077, JHEP 0211 (2002) 006.}
\lref\BB{K. Becker, M. Becker,
``M-Theory on Eight-Manifolds,''
hep-th/9605053, Nucl.Phys. B477 (1996) 155-167}
\lref\BBii{K. Becker, M. Becker, ``Supersymmetry Breaking, M-Theory and Fluxes'',
hep-th/0107044, JHEP 0107 (2001) 038.}
\lref\GKP{S. Giddings, S. Kachru, J. Polchinski,
``Hierarchies from fluxes in string compactifications'',
Phys. Rev. D 66(2002) 106006, hep-th/0105097.}
\lref\KKLT{ S. Kachru, R. Kallosh, A. Linde, S. Trivedi,
`` De Sitter vacua in string theory'', Phys. Rev. D68(2003) 046005, hep-th/0301240.}
\lref\BKQ{C. Burgess, R. Kallosh, F. Quevedo, ``de Sitter string vacua 
from supersymmetric D-terms'', JHEP 0310 (2003) 056, hep-th/0309187.}
\lref\SS{L. Susskind, ``The Anthropic Landscape of String Theory'',
hep-th/0302219.}
\lref\SSii{L. Susskind, ``Supersymmetry Breaking in the Anthropic Landscape'',   
 hep-th/0405189.}
\lref\GVW{S. Gukov, C. Vafa, E. Witten,
 ``CFT's From Calabi-Yau Four-folds'',
Nucl.Phys. B584 (2000) 69-108; Erratum-ibid. B608 (2001) 477-478}
\lref\TTi{P. Tripathy, S. Trivedi, ``Compactification with flux on K3 and Tori'',
hep-th/0301139.}
\lref\DDF{F. Denef, M. Douglas, B. Florea, ``Building a better racetrack'',
JHEP 0406(2004)034, hep-th/0404257.}
\lref\BBCQ{V. Balasubramanian, P. Berglund, J. P. Conlon, F. Quevedo,
``Systematics of Moduli Stabilisation in Calabi-Yau Flux Compactifications'',
hep-th/0502058.}
\lref\DDKG{F. Denef, M. R. Douglas, B. Florea, A. Grassi, S. Kachru,
``Fixing All Moduli in a Simple F-Theory Compactification'',
hep-th/0503124. }
\lref\AG{A. Grassi, ``Divisors on elliptic Calabi-Yau 4-folds and the superpotential in F-theory'',
alg-geom/9704008.}
\lref\W{E. Witten, ``Nonperturbative superpotentials in string theory'',
Nucl. Phys. B 474, 343 (1996), hep-th/9604030.}
\lref\SR{D. Robbins, S. Sethi, ``A Barren Landscape?'',
hep-th/0405011.}
\lref\GKTT{L. G\"orlich, S. Kachru, P. Tripathy, S. Trivedi,
`` Gaugino condensation and nonperturbative superpotentials in flux compactifications'', hep-th/0407130.}
\lref\KS{R. Kallosh, D. Sorokin,
``Dirac action on M5 and M2 branes with bulk fluxes'',
hep-th/0501081.}
\lref\TT{P. Tripathy, S. Trivedi, ``D3 Brane Action and Fermion Zero Modes in Presence of Background Flux'', hep-th/0503072.}

\Title{
  \vbox{\baselineskip12pt \hbox{hep-th/0503125}
  \hbox{}
  \vskip-.5in}
}{\vbox{
  \centerline{ Topological constraints on stabilized flux vacua
 }
\centerline{ }
  \centerline{}
}}
\centerline{ Natalia Saulina }
\bigskip\medskip
\centerline{ \it Jefferson Physical Laboratory, Harvard University, Cambridge, MA 02138, USA}
\medskip
\medskip
\medskip
\medskip
\medskip
\medskip

We study the influence of four-form fluxes on the stabilization of the K\"ahler moduli
in M-theory compactified on a
Calabi-Yau four-fold. We find that, under certain non-degeneracy condition on the flux,
M5-instantons of a new topological type generate a superpotential.
The existence
of such an instanton  restricts possible four-folds for which the
stabilization by this mechanism is expected. 
These topological constraints on the background
 are different from the previously known constraints, derived from the {\it flux-free} 
analysis of the nonperturbative effects.

\Date{March, 2005}
\newsec{Introduction}
Whether or not we wish to accept the anthropic philosophy \SS,\SSii, a necessary
condition for a plausible phenomenologically realistic background is the stabilization of all of its
moduli. In the context of the orientifold type IIB models \DRS,\GKP,\BD,\KKLT,\BKQ\
it is now clear that  the complex
structure moduli and the axion-dilaton modulus are fixed by a perturbative 
superpotential proportional to the fluxes \GVW. On the other hand, the stabilization of the K\"ahler moduli relies on the generation of the
nonperturbative superpotential. The nonperturbative effects originate
from gaugino condensation on coincident D7-branes \TTi,\DDF\
present in the background and from the D3-brane instantons
\DDF,\W,\AG,\GKTT.

The KKLT paper \KKLT\ qualitatively  discussed  the
nonperturbative superpotential deriving its intuition from the
{\it flux-free} compactifications. The subsequent successful
search for the realistic
backgrounds \DDF,\BBCQ,\DDKG\  was also based on the {\it flux-free} analysis \W,\AG\
of the nonperturbative effects. 

However, recently it was realized \GKTT,\KS,\TT\
that the presence of  background fluxes may actually modify the conditions
for the generation of an instanton-induced superpotential.

In this note we study the effect of the background flux 
on the generation of a nonperturbative superpotential
for the K\"ahler moduli. We find that, under certain restrictions on 
 the background flux, instantons of a new topological
type  generate a superpotential. 

We investigate these new instantons  in
M-theory compactified on a Calabi-Yau four-fold $CY_4$ with four-form
fluxes \BB. The
effective 3D theory has four supercharges. Moreover, if the
four-fold is elliptically fibered and the area of the elliptic fiber is
sent to zero, a new fourth dimension appears and the background is
described as a flux compactification of type IIB string theory on a
Calabi-Yau orientifold \GKP. In the framework of M-theory 
the D7-branes are described as singular fibers of the elliptic
fibration, while the D3-brane instantons become the M5-brane instantons wrapped
on the ``vertical'' divisors\foot{A vertical divisor is one that
projects to a divisor in the base B of the elliptic fibration $\pi: CY_4
\rightarrow B$. } of $CY_4$.

An M5-brane 
wrapped on a divisor in the four-fold 
generates a superpotential required for the stabilization of K\"ahler
moduli 
if there are exactly two fermionic zero modes on its
world-volume. The relevant analysis of the generalized
Dirac equation \KS\ has not yet been done in the presence of fluxes. 
The purpose of this note is to
fill in this gap. 

We find  exactly two fermionic zero modes
by restricting the choice of fluxes and global properties of the
divisors. We consider divisors with Hodge numbers 
\eqn\glob{h^{(0,1)}=0, \quad
h^{(0,3)}=0,\quad h^{(0,0)}=h^{(0,2)}=1,}
where $h^{(0,p)} $ stands for
the number of linearly independent
harmonic $(0,p)$ forms on the divisor. Our choice of the  flux is
characterized, in addition to general supersymmetry constraints
(2.3),(2.4), by  a non-degeneracy 
condition (4.6).

Note that in the absence of fluxes there would be four fermion zero modes for
divisors with these Hodge numbers and  M5-branes wrapped on such a divisor
 would not
generate a superpotential. Our result demonstrates
that the appropriate choice of the flux lifts extra fermion zero modes 
so that instantons of the previously ignored topological
type contribute to the stabilization of the K\"ahler moduli.

We would like to emphasize that it is not trivial to reduce the number of
the fermionic zero modes of the instanton to two. For example, \TT\ have
counted the number of the fermion zero modes in the context of a type IIB
compactification on the orientifold $T^6/Z_2$ in the presence of fluxes and
found four zero modes. In their case, no instanton-induced 
superpotential is generated.

The existence
of a divisor with Hodge numbers \glob\ restricts possible four-folds for which the
stabilization of the K\"ahler moduli due to M5-instantons of the new topological type
 is expected.
These topological constraints on the background
 are different from the previously known constraints derived from the {\it flux-free} 
analysis of the nonperturbative effects.

 The note is organized as follows.
In Section 2 we briefly review some basic facts about the flux
compactification of M-theory on a Calabi-Yau four-fold $CY_4$ and recall
the geometric properties of the fermions living on the M5-brane instanton.
In Section 3 we review the Dirac-like equation  for the fermions
living on the M5-brane in the presence of background fluxes. For an
M5-brane wrapped on a divisor D in $CY_4$ we recast this equation as a set
of equations for differential forms on D. In Section 4 we demonstrate that
for generic  fluxes and divisors with global properties \glob\
there are exactly two fermionic zero modes. Section 5 summarizes our 
results.

\newsec{Flux compactification of M-theory on $CY_4$ and an $M5-$instanton. } 
In this section we review basic facts about 
flux compactification of M-theory on Calabi-Yau 4-fold $CY_4$ \BB\
and recall the geometric properties of the fermions living on an M5-brane instanton.

 The 11D metric is a warped product
\eqn\metr{
ds^2=e^{2A(y)}\eta_{\mu \nu} dx^{\mu} dx^{\nu} +
e^{2B(y)}g_{MN}dy^Mdy^N}
where $\eta_{\mu \nu}$ is the metric on the three-dimensional
Minkowski space and the internal metric has the form:
\eqn\metrii{
g_{MN}=t^2g_{MN}^{(0)}+g_{MN}^{(1)}+\ldots}
Here $g_{MN}^{(0)}$ is Ricci-flat metric on $CY_4$ and $t$
is the size of the 4-fold.

In the leading approximation in the limit of large $t$ the warped factors
are trivial $A^{(0)}=B^{(0)}=0$ and the 4-form flux has
only  components along the 4-fold. Moreover, compactification gives 
3D theory with four supercharges
when the background flux is a primitive form of (2,2) type
\eqn\prim{J\wdg F_{(2,2)}=0}
and the tadpole cancellation condition is satisfied
\eqn\tad{ \int_{CY_4} F \wdg F +{\chi\over 12}=0}
In the equations \prim,\tad\ $J$ is K\"ahler form on $CY_4$ and $\chi$ is
Euler characteristic of $CY_4.$

Now let us consider a divisor D in $CY_4$ and wrap an M5-brane on it.
This M5-instanton will generate nonperturbative superpotential
$W_{np}={\it G}(Z) e^{-T}$ if fermions living on M5 brane world-volume have exactly two  zero modes. Here $T=V+iC,$ $V$ is the volume  of the divisor D
and  the axion $C$ is the 6-form potential integrated over the divisor.    
The prefactor ${\it G}(Z)$ is a 1-loop determinant which is
a holomorphic function of the complex structure moduli. 

Our goal in  Section 3 will be to recast the
equations of motion for fermions living on the M5-instanton as
a set of equations on differential forms on the divisor D.
We will further use this in  Section 4  to find the case
with exactly two fermion zero modes.

For this purpose we  recall below  how
world-volume fermions transform under the rotations of the normal and
tangent directions.
The normal bundle to the M5-brane has a product form $R^3 \times {\cal N},$ where $R^3$ stands for external space\foot{ For computation of instanton generated superpotential
we work in Euclidean signature in external 3D space. }
and ${\cal N}$ is the line bundle describing one complex normal 
direction inside $CY_4.$

The fermions $\theta=\pmatrix{\theta_{\a}^{ A +} \cr \theta_{\a}^{ A  -} }$
living on the M5-brane transform in representation 
${\bf 4 \otimes 2 \otimes 2}$ 
under $Spin(6)\times SO(3)\times SO(2)$. 
Here $A=1,2$ is a spinor under  
external $SO(3),$ the $+(-)$ stands for a chiral(anti-chiral)
spinor of SO(2)  and $\a=1,\ldots,4$  is a chiral
spinor  of Spin(6).

\newsec{Recasting equations for fermion modes of the M5-instanton
in terms of differential forms. }
R.~Kallosh and D.~Sorokin \KS\ have  derived
the Dirac-like equation for the fermions living on the
world-volume of an M5-brane in the presence of background fluxes.
We now apply their equation
to the case of the M5-brane wrapped on a divisor D of $CY_4.$
The goal of this section is to rewrite the resulting equations in terms
of differential forms on the divisor. This will simplify our
search for fermion zero modes in Section 4. The reader may skip
the details and find the resulting system of equations in
(3.6)-(3.9).

In the limit of the large size $t$ of the four-fold, all components of the four-form except those
with all indices along the internal four-fold may be neglected. The 
equation then reads:

\eqn\ferm{
{\tilde \gamma}^i \nabla_i \theta + \tilde \gamma^{\bar i} \nabla_{\bar i} \theta -
{ 1\over 8}T_{\bar w} {\tilde \gamma}^{{\bar i} {\bar j} k}F_{{\bar i} {\bar j} k}^{~~~\bar w}\theta
-{ 1\over 8}T_{w} {\tilde \gamma}^{i j {\bar k} }F_{ i j \bar{k}}^{~~~{w}}\theta=0
}
We have introduced complex coordinates $z^i,\quad i=1,2,3$ along the
divisor and the complex coordinate $w$ normal to the divisor inside $CY_4.$ 

Note that $F_{\bar i \bar j k w}$ and $F_{i j \bar k \bar w}$ are the only
internal flux components which appear in the equation \ferm. The same
components are turned on when a dual, four-dimensional type IIB
orientifold description of the three-dimensional theory becomes 
applicable. In the case of general flux compactifications of M-theory
on a Calabi-Yau four-fold there could  
also be other internal flux components\foot{$F_{ij \bar k \bar n}$ and 
$F_{i \bar j w \bar w}$.},
as long as they satisfy the supersymmetry constraints \prim,\tad.
It should be noted that remarkably, they do not affect the equation
for the instanton fermionic zero modes.

In \ferm\ $T_w,T_{\bar w} $ are $SO(2)$ Dirac matrices:
\eqn\sotwo{T_{w} T_{\bar w} + T_{\bar w} T_{w}=2g_{w \bar w}}
The six dimensional chiral(anti-chiral) gamma matrices ${ \tilde \gamma}_i^{\a \b},
\, { \tilde \gamma}_{\bar j}^{\a \b}(\gamma_{i~\a \b},\, \gamma_{{\bar j}~\a \b})$ have the properties
\eqn\sixdim{
\gamma_{\bar j}{\tilde \gamma}_i+ \gamma_{i}{\tilde \gamma}_{\bar j}=2g_{i \bar j} }
where $g_{i \bar j}$ is K\"ahler metric on the divisor D.
Note that nothing in the  equation \ferm\ acts on the
index $A=1,2$ of a spinor in $R^3.$  In what follows we will not write
this index explicitly but we will keep it in mind in the future counting
of the number of zero modes.
  
The covariant derivatives $\nabla_{j},\nabla_{\bar j}$ include 
the connection on the bundle of chiral Spin(6) spinors
 as well as connection on the spin bundle
derived from  the normal bundle ${\cal N}$.

Now we use the known fact (see for example \W) that
the bundle $S^+$ of chiral spinors on a K\"ahler manifold
of complex dimension three is isomorphic to the bundle  
$$\Bigl( \Omega^{(0,0)}\otimes K^{\half}\Bigr) \oplus
\Bigl(\Omega^{(0,2)}\otimes K^{\half}\Bigr)$$
Here $\Omega^{(0,p)}$ stands for the bundle of $(0,p)$ forms.
 We will further use that 
the normal bundle on the divisor in $CY_4$
is isomorphic to the canonical bundle $K.$ 
Recalling that  $\theta$ is a section  
 of the bundle\foot{Here we are ignoring that $\theta$ is a spinor  in $R^3$.} $S^+\otimes K^{\half} \oplus S^+\otimes K^{-\half},$
we find the following degrees of freedom.   
A (0,2) form $a_{(2)}^w$ taking values in the canonical bundle K,
a section of K  $a_{(0)}^w $ as well as
a (0,2) form $b_{(2)}$
and a  scalar $b_{(0)}. $

Locally we write $\theta$ in terms of these degrees of freedom
as follows 
 \eqn\forms{ \theta=\bigl( a_{(0)}^w +a_{{\bar i} {\bar j}}^w\gamma^{\bar i} {\tilde \gamma}^{\bar j}\bigr)T_{w}\e+ 
 \bigl( b_{(0)} +b_{{\bar i} {\bar j}}\gamma^{\bar i} 
{\tilde \gamma}^{\bar j}\bigr)\e}
where the  chiral spinor $\e$ satisfies 
\eqn\ugu{{\tilde \gamma}^i\e=0,\quad i=1,2,3,\quad T_{\bar w} \e=0}

Plugging \forms\ into \ferm\
we find the following set of equations\foot{
$X_{[\bar i_1 \ldots \bar i_p]}={1 \over p!}\bigl( X_{\bar i_1 \ldots \bar i_p} \pm permutations\bigr)$}:
\eqn\geom{\p_{[\bar i} b_{{\bar j}{\bar k}]}=0}
\eqn\geomi{ 4 \p^{\bar j}b_{\bar j \bar k}+\p_{\bar k} b_{(0)}=0  }
\eqn\geomii{ D_{[\bar m_1} a_{\bar m_2 \bar m_3]}^w=0 }
\eqn\geomiii{4D^{\bar j} a_{\bar j \bar k}^w+
 D_{\bar k}a_{(0)}^w=-F^{{\bar i} {\bar j}~~w}_{~~~\bar k } b_{\bar i \bar j}}
In the equations \geomii,\geomiii\ the covariant
differentials include  connection on the canonical
bundle.
In writing the equations \geom-\geomiii\  we used the primitivity condition \prim.

\newsec{The M5-instanton with two fermion zero modes.}

Here we study the set of equations \geom-\geomiii\ for the
fermionic degrees of freedom on an M5-brane wrapped on a divisor D in a Calabi-Yau 4-fold.
We  consider divisors with Hodge numbers
\eqn\prop{ h^{(0,1)}=h^{(0,3)}=0,\quad h^{(0,0)}=h^{(0,2)}=1}
where $h^{(0,p)} $ stands for the number of harmonic $(0,p)$ forms on D.
The goal of this section is to show that for generic 
background fluxes the M5-branes wrapped on the 
 divisors of this topological type 
generate a superpotential. 
 
In the absence of fluxes there would be four fermion zero modes
for the divisors with these Hodge numbers. Two zero modes\foot{Recalling that all fields carry 
   spinor  index in $R^3$.} would be coming from
harmonic (0,2) form and the other two from (0,0) form. It is natural to expect
that choosing flux appropriately one can  lift zero modes associated
with (0,2) form. Below we  realize this expectation.

Using Hodge decomposition the equations \geom\ and \geomi\ 
imply that $b_{(2)} $ and $b_{(0)}$ are harmonic forms.
From $h^{(0,2)}=1$ follows that we may  write $b_{\bar i \bar j}=\b \omega_{\bar i \bar j}$
 where $\omega_{\bar i \bar j}$ is a fixed harmonic (0,2) form
and $\b$ is a complex number.

Now let us consider equation \geomiii. Both sides of this equation
take values in $\Omega^{(0,1)}(K),$ the space of (0,1)-forms with values in the canonical 
bundle K. 
From our assumption $h^{(0,2)}=1$ follows $h^{(0,1)}(K)=1.$ This implies that there is unique (up to multiplication
by complex number) harmonic 
(0,1)-form taking values in K. Let us call it $c_{\bar k}^w.$  
Now we take inner product
of both sides of \geomiii\ with $ c_{\bar k}^w.$ 
The left side gives zero. So the consistency of \geomiii\ requires 

\eqn\cons{ \int_D \sqrt{g}\, g_{w \bar w}
\bigl( c_{\bar k}^w\bigr)^* g^{k \bar p} F^{\bar l \bar m ~~w}_{~~~\bar p} b_{\bar l \bar m}=0}

Recall also that $c_{\bar k}^w$ can be constructed from \foot{This is
explicit realization of the statement $h^{(0,1)}(K)=h^{(0,2)}.$} 
the fixed harmonic (2,0) form $\omega_{i j}=(\omega_{\bar i \bar j})^*$ as follows:
\eqn\constr{ c_{\bar k}^w=g_{\bar k p} \varepsilon^{pijw}\omega_{ij} }
where $\varepsilon^{pijw}$ is $SU(4)$ invariant anti-symmetric tensor.

The equation \cons\ becomes
\eqn\consi{\b \int_D \sqrt{g}\, \omega_{\bar i \bar j} P^{\bar i \bar j, \bar k \bar m}\omega_{\bar k \bar m}=0 } 
where
\eqn\defp{P^{\bar i \bar j, \bar k \bar m}=g_{w \bar w} \varepsilon^{\bar p \bar i \bar j \bar w}F^{\bar k \bar m~~w} _{~~~~\bar p}}
So we conclude that for generic fluxes such that
\eqn\choice{ \int_D \sqrt{g} \omega_{\bar i \bar j} P^{\bar i \bar j, \bar k \bar m}\omega_{\bar k \bar m} \ne 0}
the only solution of \consi\ is $\b=0$ and therefore $b_{(2)}=0.$
We would like to emphasize that this removal of the harmonic (0,2)-form
$b_{(2)}$ is eventually responsible for lifting of the 
two extra fermion zero modes. 

Now equations \geomii\ and \geomiii\ require 
$a_{(0)}^w $ and $a_{(2)}^w$ to be harmonic forms
with values in K. From our assumption about the topology
of the divisor D $h^{(0,3)}=h^{(0,1)}=0$
we find $h^{(0,0)}(K)=h^{(0,2)}(K)=0$ and therefore
$a_{(0)}^w=0 $ and $a_{(2)}^w=0.$

We conclude that the
 equations \geom-\geomiii\   have a single solution:
$$ b_{(0)}=const, \quad a_{(0)}^w=0,\quad a_{(2)}^w=0,\quad b_{(2)}=0$$

Recalling that all the fields carry hidden  index $A=1,2$ 
of a spinor in $R^3$ (see discussion below \sixdim), we  conclude
that we found exactly two fermion zero modes.  Therefore,  the M5-instanton
of the topology \prop\ in the presence of generic fluxes \choice\ 
generates nonperturbative
superpotential for the K\"ahler moduli.  

\newsec{Conclusion}
In this note we studied how the conditions for the 
stabilization of the K\"ahler moduli 
are modified by background fluxes. 
We considered M-theory compactified on a Calabi-Yau
four-fold 
and found that, for a generic choice of background fluxes, 
M5-instantons of a new topological type generate a nonperturbative superpotential
 required for the stabilization.

The new instanton  is an M5-brane wrapped on a divisor
 with   Hodge numbers
$$h^{(0,0)}=h^{(0,2)}=1, \quad h^{(0,1)}=h^{(0,3)}=0$$
Meanwhile, the background fluxes, in addition to general 
constraints on supersymmetric compactification \prim,\tad,
are characterized by 
a non-degeneracy condition \choice.

Divisors with these Hodge numbers appeared  before\foot{The example in \GKTT\ is $K_3\times P^1$
in $K_3\times K_3.$} in
the discussion of the gaugino condensation on coincident D7-btanes \GKTT.
We found that in the presence of the special background fluxes 
such divisors are relevant for 
the generation of nonperturbative superpotential induced by
the M5-instantons.
  
The condition for the existence
of such a divisor restricts possible four-folds for which the
stabilization of the K\"ahler moduli by this mechanism is expected.
These  topological constraints on the background
 are different from the previously known constraints derived from the {\it flux-free} 
analysis of the nonperturbative effects.

It would be interesting to find other choices of fluxes which can make the
M5-branes wrapped on more general divisors to contribute to
the nonperturbative superpotential. Another interesting question is to find
an explicit example of a Calabi-Yau four-fold, other than $K_3\times K_3,$ that  contains a divisor with the desired properties \prop\ and admits the appropriate 
non-degenerate flux \choice.
\medskip
\medskip
{\bf Acknowledgments}

I would like to thank L. Motl and J.Distler  for  valuable discussions. 
My research was supported in part
by NSF grants PHY-0244821 and DMS-0244464.   
I am grateful to the referee of Nuclear Physics B who pointed
out a mistake in the original version of this paper.

\listrefs
\end